\documentclass[manuscript]{acmart}

\usepackage{amsmath}
\usepackage{cleveref}
\usepackage{pifont}
\usepackage{booktabs}
\usepackage{algorithm}
\usepackage[noend]{algpseudocode}
\usepackage{multirow}
\usepackage{caption}
\usepackage{subcaption}

\newcommand{\change}[1]{\textcolor{black}{#1}}

\AtBeginDocument{%
  }

\copyrightyear{2022} 
\acmYear{2022} 
\setcopyright{rightsretained} 
\acmConference[RecSys '22]{Sixteenth ACM Conference on Recommender Systems}{September 18--23, 2022}{Seattle, WA, USA}
\acmBooktitle{Sixteenth ACM Conference on Recommender Systems (RecSys '22), September 18--23, 2022, Seattle, WA, USA}
\acmDOI{10.1145/3523227.3546755}
\acmISBN{978-1-4503-9278-5/22/09}

\begin{document}

\title{Bundle MCR: Towards Conversational Bundle Recommendation}


\author{Zhankui He}
\email{zhh004@eng.ucsd.edu}
\affiliation{%
  \institution{UC San Diego}
  \city{San Diego}
  \state{California}
  \country{USA}
}

\author{Handong Zhao}
\authornote{The corresponding author.}
\email{hazhao@adobe.com}
\affiliation{%
  \institution{Adobe Research}
  \city{San Jose}
  \state{California}
  \country{USA}
 }

\author{Tong Yu}
\email{tyu@adobe.com}
\affiliation{%
  \institution{Adobe Research}
  \city{San Jose}
  \state{California}
  \country{USA}
 }

\author{Sungchul Kim}
\email{sukim@adobe.com}
\affiliation{%
  \institution{Adobe Research}
  \city{San Jose}
  \state{California}
  \country{USA}
}

\author{Fan Du}
\email{fdu@adobe.com}
\affiliation{%
  \institution{Adobe Research}
  \city{San Jose}
  \state{California}
  \country{USA}
 }

\author{Julian McAuley}
\email{jmcauley@eng.ucsd.edu}
\affiliation{%
  \institution{UC San Diego}
  \city{San Diego}
  \state{California}
  \country{USA}
}

\renewcommand{\shortauthors}{He et al.}

\begin{abstract}
    \emph{Bundle} recommender systems recommend sets of items (e.g.,~pants, shirt, and shoes) to users, but they often suffer from two issues: significant \emph{interaction sparsity} and a \emph{large output space}.
    In this work, we extend \emph{multi-round conversational recommendation} (MCR) to alleviate these issues. MCR---which uses a conversational paradigm to elicit user interests by asking user preferences on tags (e.g.,~categories or attributes) and handling user feedback across multiple rounds---is an emerging recommendation setting to acquire user feedback and narrow down the output space, but has not been explored in the context of bundle recommendation. 
    
    In this work, we propose a novel recommendation task named \textit{Bundle MCR}. Unlike traditional bundle recommendation (a bundle-aware user model and bundle generation), Bundle MCR  studies how to encode user feedback as conversation states and how to post questions to users. Unlike existing MCR in which agents recommend individual items only, Bundle MCR handles more complicated user feedback on multiple items and related tags. To support this, we first propose a new framework to formulate Bundle MCR as Markov Decision Processes (MDPs) with multiple agents, for user modeling, consultation and feedback handling in bundle contexts. Under this framework, we propose a model architecture, called Bundle Bert (\textsc{Bunt}) to \textit{(1)~recommend items}, \textit{(2)~post questions} and  \textit{(3) manage conversations} based on bundle-aware conversation states. Moreover, to train \textsc{Bunt} effectively, we propose a two-stage training strategy. In an offline pre-training stage, \textsc{Bunt} is trained using multiple \textit{cloze} tasks to mimic bundle interactions in conversations. Then in an online fine-tuning stage, \textsc{Bunt} agents are enhanced by user interactions. Our experiments on multiple offline datasets as well as the human evaluation show the value of extending MCR frameworks to bundle settings and the effectiveness of our \textsc{Bunt} design.
\end{abstract}

\ccsdesc[500]{Information systems~Personalization}

\keywords{recommender systems, conversational recommendation, bundle recommendation}

\maketitle

\section{Introduction}

\begin{figure*}[t]
  \centering
  \includegraphics[width=\linewidth]{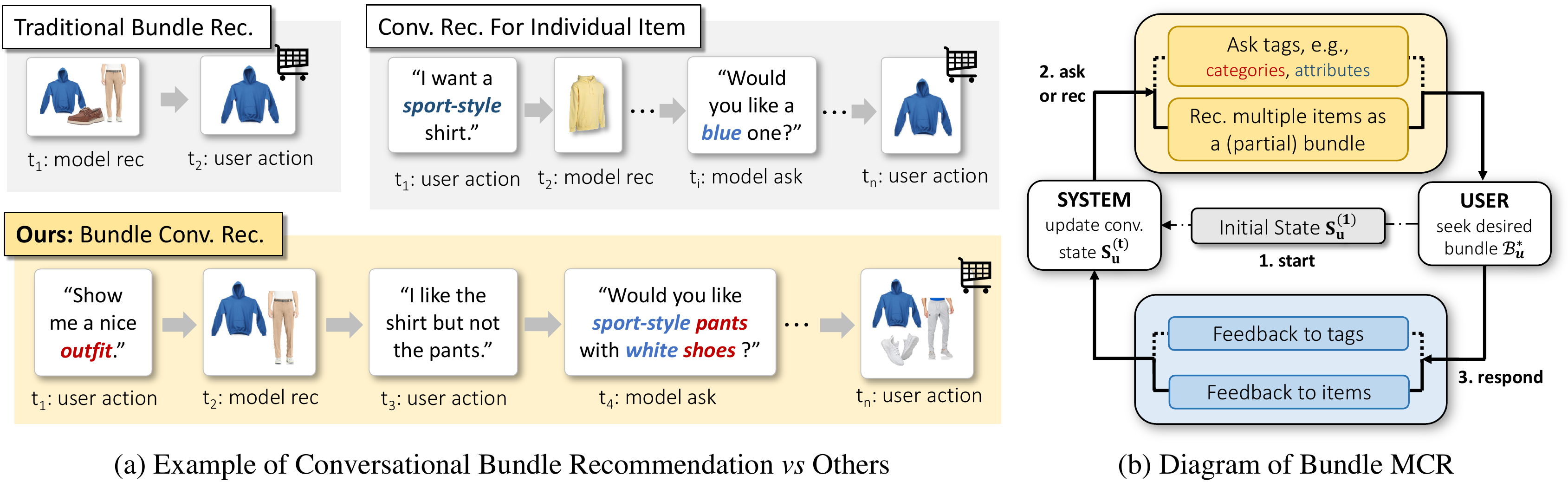}
  \caption{\emph{Left}:  Use case comparison among traditional bundle recommendation, individual conversational recommendation and our conversational bundle recommendation. \emph{Right}: Diagram of our proposed Bundle-Aware Multi-Round Conversational Recommendation (i.e.,~Bundle MCR) scenario, which extends traditional individual MCR~\cite{Lei2020EstimationActionReflectionTD, Lei2020InteractivePR, UNICORN, MIMCR} to bundle setting.}
  \label{fig:intro}
\end{figure*}

Bundle recommendation aims at recommending sets of items that can be simultaneously consumed by users~\cite{Pathak2017GeneratingAP, Chen2019MatchingUW, Deng2021BuildYO} (e.g.,~outfits, playlists), which improves user satisfaction~\cite{Deng2020PersonalizedBR, Chen2019POGPO}. However, bundle recommendation is inherently challenging due to (at least) two issues: (1)~\textbf{Interaction sparsity}: user-bundle interactions are much sparser than user-item interactions, leading to the difficulty of modeling user preferences accurately;  (2)~\textbf{Output space complexity}: predicting a correct bundle (i.e.,~multiple items) from all item combinations is more challenging than traditional individual item recommendations.

Currently, two approaches are proposed in bundle recommendation to circumvent these issues. The first line~\cite{Pathak2017GeneratingAP, Chang2020BundleRW, Chen2019MatchingUW} presents \emph{discriminative} methods, i.e.,~ranking \emph{existing} bundles which avoids the complexity issue, by treating each bundle as a generalized individual `item'.
The application scenarios of those methods are usually narrow (e.g.,~for pre-defined bundle sales). The second line~\cite{Bai2019PersonalizedBL, Deng2021BuildYO, Hu2019Sets2SetsLF} uses \emph{generative} methods, i.e.,~generating 
(perhaps new) 
bundles, which is more flexible but still suffers from limited accuracy. In these works, bundle recommenders are \emph{one-shot}, i.e.,~recommending a complete bundle with a single attempt. As the traditional bundle recommendation in~\Cref{fig:intro}a shows, the user receives a completed bundle (shirt, shoes and pants) and reacts to this bundle (picking a shirt but ignoring others), then recommendation ends. Clearly, such one-shot setting doesn't allow the model to collect continuous user feedback and provide enhanced bundle with higher accuracy. Considering such limitations, we present a new \emph{multi-round} and \emph{interactive} way for bundle recommendation, i.e.,~allowing the user and system to ``discuss'' bundle composition together. Specifically, we call this \emph{multi-round conversational bundle recommendation} task \emph{Bundle MCR}.

The core idea of Bundle MCR is to extend one of the representative conversational recommendation mechanisms -- multi-round conversational recommendation (MCR)~\cite{Lei2020EstimationActionReflectionTD, Lei2020InteractivePR, UNICORN, MIMCR} -- to bundle contexts, in which the system can acquire user feedback on item tags and narrow down the output space during conversations for more accurate bundle recommendation. Although recently many conversational recommendation works~\cite{Christakopoulou2016TowardsCR, Lei2020EstimationActionReflectionTD, Zhang2020ConversationalCB,Lei2020InteractivePR, UNICORN, MIMCR}, especially MCR frameworks~\cite{Lei2020EstimationActionReflectionTD, Lei2020InteractivePR, UNICORN, MIMCR} have proven effective to elicit user preferences for individual item recommendation, designing a new MCR framework for bundle recommendation is still non-trivial: existing MCR frameworks target recommending an individual item only (named Individual MCR); they cannot directly work for bundle settings for several reasons: (1)~not considering user-bundle interactions or item-item relationships in user preference modeling; (2)~recommending top-K individual items instead of multiple items as a bundle (or partial bundle); (3)~handling user feedback and posting questions on tags related to an individual target item, without considering feedback or questions to different items within a bundle. We illustrate the gap between Individual MCR and Bundle MCR in~\Cref{fig:intro}a. Individual MCR updates user feedback on tags (e.g.,~attributes like \textit{sport-style}, \textit{blue}) to narrow down the candidate item pool effectively but cannot post questions or model the feedback to multiple items (i.e., bundle-aware) directly. Instead, Bundle MCR aims to complete a bundle with the user by generating multiple items as a bundle or partial bundle, and handling questions to multiple items (e.g., \textit{sport-style} \textit{pants} and \textit{white} \textit{shoes}).

Methodologically, we formulate Bundle MCR as a Markov Decision Process (MDP) problem with multiple agents. Then, we propose a new model architecture Bundle Bert (\textsc{Bunt}) to conduct these functions in a unified self-attentive~\cite{SelfAtt, Kang2018SelfAttentiveSR, Sun2019BERT4RecSR} architecture. Furthermore, to train \textsc{Bunt} efficiently, a two-stage training strategy is proposed: we pre-train \textsc{Bunt} with multiple \textit{cloze} tasks, to learn the basic knowledge of how to infer correct items, tags and when to ask or recommend based on conversation contexts mimicked by offline user-bundle interactions. Then, we introduce a user simulator, create a simulated online environment, and fine-tune \textsc{Bunt} agents with reinforcement learning on conversational bundle interactions with users. We summarize the contributions of this work as below:

\begin{itemize}
    \item We propose a Bundle MCR setting where users and the system complete a bundle together. To our knowledge, this is the first work that considers a conversational mechanism in bundle recommendation and also alleviates the bundle recommendation issues of \emph{information sparsity} and \emph{output space complexity}.
    \item We present an MDP framework with multiple agents for Bundle MCR. Under this framework, we propose Bundle Bert (\textsc{Bunt}) to conduct multiple Bundle MCR functions in a unified self-attentive architecture. We also design a two-stage (pre-training and fine-tuning) strategy for \textsc{Bunt} learning.
    \item We evaluate conversational bundle recommendations on four offline bundle datasets and conduct a human evaluation, to show the effectiveness of \textsc{Bunt} and the potential of conversational bundle recommendation.
\end{itemize}

\begin{table}[]
\small
\caption{Functionality requirements in Bundle MCR model design and comparisons with individual MCR and bundle recommendation.}
\begin{tabular}{lccc}
\toprule
\textbf{Functionalities  }                 & \textbf{Individual MCR} & \textbf{Bundle Recommendation} & \textbf{Bundle MCR (ours)} \\ \midrule
Bundle-Aware User Modeling    &          \ding{56}               &           \cite{Deng2021BuildYO, Bai2019PersonalizedBL, Pathak2017GeneratingAP, Chang2020BundleRW, Chen2019MatchingUW, Chen2019POGPO}                     &        \ding{52}             \\
Bundle Generation             &          \ding{56}               &           \cite{Deng2021BuildYO, Bai2019PersonalizedBL, Pathak2017GeneratingAP, Chen2019POGPO}                     &        \ding{52}             \\
Bundle-Aware Feedback Handing &          \ding{56}               &           \ding{56}                     &        \ding{52}             \\
Bundle-Aware Question Asking  &          \ding{56}               &           \ding{56}                     &        \ding{52}             \\
Conversation Management       &          \cite{Lei2020EstimationActionReflectionTD, Lei2020InteractivePR, Zhang2020ConversationalCB,UNICORN}               &           \ding{56}                     &        \ding{52}             \\ \bottomrule
\end{tabular}
\label{tab:comp}
\end{table}

\section{Related Work}
\label{sec:related}

\subsection{Bundle Recommendation}

Bundle (or set, basket) recommendation offers multiple items as a set to user. Traditional bundle recommendation adopts Integer Programming~\cite{MarchettiSpaccamela1995StochasticOK, Xie2010BreakingOO, zhu2014bundle} or Association Analysis~\cite{Chen2005MarketBA, Garfinkel2006DesignOA} Most of them have no personalization. \change{Some works~\cite{Wibowo2018IncorporatingCI, zanker2010constraint} apply constraint solvers}. Recently, more works are learning-based and can be divided into two categories: \textbf{(1)~discriminative methods:} Bundle BPR (BBPR)~\cite{Pathak2017GeneratingAP} extends BPR~\cite{Rendle2009BPRBP} to personalized bundle ranking (BBPR also designed a heuristic generative algorithm but it is time-consuming); DAM~\cite{Chen2019MatchingUW} and BGCN~\cite{chang2020bundle} enhance the representation of users with factorized attention networks or graph neural networks. \textbf{(2)~generative methods:} An encoder-decoder framework is used in BGN~\cite{Bai2019PersonalizedBL} (RNN~\cite{Cho2014OnTP}-based) and PoG~\cite{Chen2019POGPO} (Transformer~\cite{SelfAtt}-based) to generate multiple items as a personalized bundle. BGN decomposes bundle recommendation into quality/diversity via determinantal point processes. BYOB~\cite{Deng2021BuildYO} treats bundle generation as a sequential decision making problem with reinforcement learning methods. In our work, the bundle recommender is for interactive (conversational) settings. As~\Cref{tab:comp} shows, existing bundle recommenders are \emph{one-shot} so they focus on user modeling and bundle generation only. Our model (\textsc{Bunt}) also considers how to handle user feedback, post questions and manage conversations in a unified architecture.

\subsection{Conversational Recommendation}

Conversational recommender system (CRS) enables systems to converse with users actively. CRSs seek to ask questions (e.g.~`which one do you prefer') to establish user preferences efficiently or to explain recommendations. Existing CRS methods can be classified by the question spaces:
\textbf{(1)~Asking free text:} this method generates human-like responses in natural language~\cite{Kang2019RecommendationAA, Li2018TowardsDC, Chen2019TowardsKR}. For example, \cite{Li2018TowardsDC} collects a natural-language conversational recommendation dataset \emph{ReDial} and builds a hierarchical RNN framework on it. KBRD~\cite{Chen2019TowardsKR} further incorporates knowledge-grounded information to unify recommender systems with dialog systems. \textbf{(2)~Asking about items:}~\cite{Christakopoulou2016TowardsCR, Xie2021ComparisonbasedCR} For example, \cite{Christakopoulou2016TowardsCR} designs absolute (i.e.,~want item \emph{A}?) or relative-question templates (i.e.~item \emph{A} or \emph{B}?) and evaluates several question strategies such as \emph{Greedy}, \emph{UCB}~\cite{Auer2002UsingCB} or \emph{Thompson Sampling}~\cite{Chapelle2011AnEE}; \textbf{(3)~Asking about tags:} the system is allowed to ask questions on user preference over different tags associated with items. For example, CRM~\cite{Sun2018ConversationalRS} integrates conversation and recommender systems into a unified deep reinforcement learning framework to ask facets (e.g.~color, branch) and recommend items. SAUR~\cite{Zhang2018TowardsCS} proposes a \emph{System Ask-User Respond} paradigm to ask pre-defined questions about item attributes in the right order and provide ranked lists to users. The multi-round conversational recommendation (MCR)~\cite{Lei2020EstimationActionReflectionTD, Lei2020InteractivePR, UNICORN, MIMCR} setting also belongs to conversational recommendation setting (3).

\subsection{Multi-Round Conversational Recommendation (MCR)}
\label{sec:MCR}

In our work, we focus on MCR setting, based on the following logic: (1)~Completing a bundle is naturally a multi-round process, in which more user feedback to item tags is collected to make more accurate recommendations and put more items into the potential bundle. (2)~MCR is arguably the most realistic setting 
available~\cite{Lei2020EstimationActionReflectionTD, Lei2020InteractivePR, UNICORN, MIMCR} and widely used in recent conversational recommenders. For example, EAR~\cite{Lei2020EstimationActionReflectionTD} proposes a \emph{Estimation-Action-Reflection} framework to ask attributes and model users' online feedback. Furthermore, SCPR~\cite{Lei2020InteractivePR} incorporates an item-attribute graph to provide explainable conversational recommendations. UNICORN~\cite{UNICORN} proposes a unified reinforcement learning framework based on dynamic weighted graph for MCR. To make individual MCR more realistic, MIMCR~\cite{MIMCR} allows users in MCR to select multiple choices for questions, and model user preferences with multi-interest encoders. However, existing MCR frameworks are proposed for individual item recommendation (i.e., Individual MCR). Thus the entire model architecture (e.g., FM~\cite{Rendle2010FactorizationM}) and question strategy design is not compatible with bundle contexts. As~\Cref{tab:comp} shows, our work uses a similar conversation management idea as exiting individual MCRs, but we design model architectures for bundle-aware user modeling, question asking, feedback handling and bundle generation. 

\section{Bundle MCR Scenario}
\label{sec:setting}

We extend multi-round conversational recommendation (MCR)~\cite{Lei2020EstimationActionReflectionTD, Lei2020InteractivePR, UNICORN} to a bundle setting (i.e.,~Bundle MCR). Different from individual MCR, we propose a new concept \emph{slot} for bundle MCR\footnote{This is not the \emph{slot} concept in dialog systems.}, i.e.,~the placeholder for a consulted item. For example, an outfit (1: shoes, 2: pants, 3: shirt) has three \emph{slots} $\mathcal{X}=\{1, 2, 3\}$. Ideally, bundle MCR (1)~determines the number of slots; (2)~fills target items in the slots during conversations. 

Bundle MCR is formulated as: given the set of users $\mathcal{U}$ and items $\mathcal{I}$, we collect tags corresponding to items, such as the set of attributes $\mathcal{P}$ (e.g.,~``dark color'') and categories $\mathcal{Q}$ (e.g.,~``shoes'').  As illustrated in~\Cref{fig:intro}b, for a user $u\in\mathcal{U}$:
\begin{enumerate}
    \item Conversation starts from a state $\mathbf{S}^{(1)}_{u}$ which encodes user historical interactions $\{\mathcal{B}_1, \mathcal{B}_2, \dots\}$, where $\mathcal{B}_*$ represents a bundle of multiple items. Let us set conversational round $t=1$, the system creates multiple slots as $\mathcal{X}^{(t)}$.
    \item Then, the system decides to recommend or ask, i.e.,~(i)~recommending $|\mathcal{X}^{(t)}|$ items as a (partial) bundle to fill these proposed slots, denoting as $\mathcal{B}^{(t)}_u = \{\hat{i}_x\mid x\in \mathcal{X}^{(t)}\}$; or (ii)~asking for user preference per slot on attributes $\mathcal{A}^{(t)}_u=\{\hat{a}_x\mid x\in \mathcal{X}^{(t)}\}$ and categories $\mathcal{C}^{(t)}_u=\{\hat{c}_x\mid x\in \mathcal{X}^{(t)}\}$. Here in each slot $x$, $\hat{i}_x\in\mathcal{I}$, $\hat{a}_x\in\mathcal{P}$ and $\hat{c}_x\in\mathcal{Q}$.
    \item Next, user $u$ is required to provide feedback (i.e., accept, ignore, reject) to the proposed partial bundle $\mathcal{B}^{(t)}_u$ or attributes $\mathcal{A}^{(t)}_u$ and categories $\mathcal{C}^{(t)}_u$ per slot $x\in \mathcal{X}^{(t)}$. 
    \item After that, the system updates user feedback into new state $\mathbf{S}^{(t+1)}_{u}$, records all the accepted items into a set, denoting as $\check{\mathcal{B}}_u$\footnote{We use the $\backslash\texttt{check}$ mark for the meaning of ``being accepted by user''; similarly, we use $\backslash\texttt{hat}$ for the meaning of ``being proposed to user''.}, and updates the slots of interest as $\mathcal{X}^{(t+1)}$ by creating new slots and removing the slots $x$ in which user has accepted the recommended item $\hat{i}_x$.
\end{enumerate}
After multiple rounds of step (2)-(4), the system collects rich contextual information and create bundle $\mathcal{\check{B}}_u$ for user. The conversation terminates when $u$ is satisfied with the current bundle (i.e.,~$\check{{\mathcal{B}}}_u$ equals the target bundle $\mathcal{B}_u^*$) or this conversation reaches the maximum number of rounds $T$. 

In Bundle MCR, we identify several interesting questions: (1)~how to encode user feedback to bundle-aware state $\mathbf{S}_u^{(t+1)}$? (2)~how to accurately predict bundle-aware items or tags? (3)~how to effectively train models in Bundle MCR? (4)~how to decide the size of slots $\mathcal{X}^{(t)}$ per round? In this work, we focus on (1)-(3), and use a simple strategy for (4), i.e.,~keeping the size of slots as a fixed number $K$. Though the slot size per round is fixed, the final bundle sizes are diverse due to 
different user feedback and conversation rounds. We leave more flexible slot strategies for future works.

Note that we use attribute set $\mathcal{P}$ and category set $\mathcal{Q}$ and related models for all baselines and proposed methods. But for ease of description, we only take the attribute set $\mathcal{P}$ as the example of tags in following methodology sections.
\section{General Framework}
\label{sec:framework}

We formulate Bundle MCR as a two-step Markov Decision Process (MDPs) problem with multiple agents, since (1)~the system makes two-step decisions for first recommending or asking (i.e., conversation management), then what to recommend or ask; (2)~multiple agents are responsible for different decisions: an agent (using $\pi_M$) is for conversation management; a bundle agent (using $\pi_I$) decides to recommend items to compose a bundle; an attribute agent (using $\pi_A$) considers which attributes to ask. The goal of our framework is to maximize the expected cumulative rewards to learn different policy networks $\pi^*_M, \pi^*_I, \pi^*_A$. We divide a conversation round into user modeling, consultation, and feedback handling like~\cite{MIMCR}, then we describe our \textit{state}, \textit{policy}, \textit{action} and \textit{transition} design under this framework in related stages.

\subsection{States: Bundle-Aware User Modeling} 
\label{sec:state}

We first introduce the shared conversation state $\mathbf{S}^{(t)}_u$ for all agents. $\mathbf{S}^{(t)}_u$ is encoded (specific encoder is introduced in~\Cref{sec:model}) from the conversational information $\mathbb{S}^{(t)}_u$ at conversational round $t$, which is defined as:
\begin{equation}
    \mathbb{S}_u^{(t)} = \big(\underbrace{\vphantom{\{(\check{i}_x^{(t)},  \check{\mathcal{A}}^{(t)}_x)\mid x\in \mathcal{X}^{(\le t)}\}} {\{\mathcal{B}_1, \mathcal{B}_{2}, \dots\},}}_{\text{\small long-term preference}}\enspace 
    \underbrace{\{(\check{i}_x^{(t)},  \check{\mathcal{A}}^{(t)}_x)\mid x\in \mathcal{X}^{(\le t)}\},}_{\text{\small short-term contexts}}\quad
    \underbrace{\vphantom{{\{(\check{i}_x^{(t)},  \check{\mathcal{A}}^{(t)}_x)\mid x\in \mathcal{X}^{(\le t)}\}}} {\{(\mathcal{I}_x^{(t)}, \mathcal{P}_x^{(t)}) \mid x\in \mathcal{X}^{(\le t)}\} } }_{\text{\small candidate pools}}\enspace\big).
\end{equation}
\begin{itemize}
    \item \textbf{Long-term preference} is represented by the set of user $u$'s historical bundle interactions $\{\mathcal{B}_1, \mathcal{B}_2, \dots\}$.
    \item \textbf{Short-term contexts} collect accepted items and attributes in conversations before conversational round $t$. $\mathcal{X}^{(\le t)}$ is the set of slots till rounds $t$, i.e., $\mathcal{X}^{(\le t)}=\bigcup^{t}_{t'=1}\mathcal{X}^{(t')}$. In slot $x$ at round $t$, we record the tuple $(\check{i}_x^{(t)}, \check{\mathcal{A}}^{(t)}_x)$, where $\check{i}_x^{(t)}$ denotes the item id accepted by the user. If no item accepted in slot $x$, $\check{i}_x^{(t)}$ is set as a mask token~\texttt{[MASK]}; $\check{{\mathcal{A}}}^{(t)}_x$ is the set of accepted attributes in slot $x$. For example, the initial short-term context is a $K$-sized set of $(\texttt{[MASK]}, \emptyset)$ tuples, meaning we know nothing about accepted items or attributes.\footnote{We can record rejected items or attributes as well, but we omit them since they are currently not effective empirically in our experiments.}
    \item \textbf{Candidate pools} contain item and attribute candidates per slot at round $t$ (it is not space costly by black lists). They are initialed as completed pools $\mathcal{I}$ and $\mathcal{P}$, and updated with user $u$'s feedback,  described in~\Cref{sec:transition}. 
\end{itemize}
Second, we introduce an additional conversation information $\mathbb{\bar{S}}_u^{(t)}$ (encoded as additional state $\mathbf{\bar{S}}_u^{(t)}$ in~\Cref{sec:bunt-user-modeling}) for conversation management agent. $\mathbb{\bar{S}}_u^{(t)}$ records the result id of previous t-1 rounds as a list, such as [rec\_fail, ask\_fail, ...]. It is a commonly used state representation for conversation management agent in~\cite{Lei2020EstimationActionReflectionTD, Lei2020InteractivePR, UNICORN}. We follow the result id settings as~\cite{Lei2020EstimationActionReflectionTD}, but apart from ``rec\_suc'' id for successfully recommending a single item, we further introduce a ``bundle\_suc'' id to record the result of successfully recommending the entire bundle.

\subsection{Policies and Actions: Bundle-Aware Consultation}
\label{sec:policy}

The system moves to the consultation stage after getting conversation states in user modeling stage. Now, the system makes a two-step decision: (1)~whether to recommend or ask (using policy $\pi_M$); (2)~what to recommend (using policy $\pi_I$) or what to ask (using policy $\pi_A$). We define these policies as:
\begin{itemize}
    \item \textbf{$\pi_M$ -- conversation management:} use $\mathbf{\bar{S}}_u^{(t)}$ and $\mathbf{S}_u^{(t)}$ to predict a binary action (recommending or asking).
    \item \textbf{$\pi_I$ -- (partial) bundle generation:} if recommending, the agent uses $\mathbf{S}_u^{(t)}$ as input to generate $|\mathcal{X}^{(t)}|$ (i.e.,~$K$) items as  $\mathcal{B}^{(t)}_u = \{\hat{i}_x\mid x\in \mathcal{X}^{(t)}\}$,  where $\hat{i}_x$ is the action corresponding to slot $x$ and the actions space is $\mathcal{I}_x^{(t)}$.
    \item \textbf{$\pi_A$ -- attributes consultation:} if asking, the agent uses $\mathbf{S}_u^{(t)}$ as input to generate $|\mathcal{X}^{(t)}|$ (i.e.,~$K$) attributes as  $\mathcal{A}^{(t)}_u = \{\hat{a}_x\mid x\in \mathcal{X}^{(t)}\}$,  where $\hat{a}_x$ is the action corresponding to slot $x$ and the actions space is $\mathcal{P}_x^{(t)}$.
\end{itemize}

\subsection{Transitions: Bundle-Aware Feedback Handling} 
\label{sec:transition}

The system handles user feedback in a transition step. The user $u$ will react to the proposed $K$ items or attributes with acceptance, rejection or ignoring. Generally, in our transition step, ``acceptance'' is mainly used to update short-term contexts, ``rejection'' is used to update candidate pools and we change nothing when getting ``ignoring''. 
\begin{itemize}
    \item \textbf{Update $\mathbb{S}^{(t+1)}_u$:} long-term preference is fixed, we update the short-term contexts and candidate pools as follows:
    \textbf{(1)~Feedback to items:} for each consulted item $\hat{i}_x$, (i) all item candidates pools $\mathcal{I}^{(t+1)}_{x'}$ where $x'\in \mathcal{X}^{(t)}$ delete $\hat{i}_x$ because it has been recommended; (ii) if $\hat{i}_x$ is accepted, short-term contexts in slot $x$ will assign $\hat{i}_x$ to $\check{i}^{(t)}_x$.
    \textbf{(2)~Feedback to attributes:} for each consulted attribute $\hat{a}_x$, (i)~different from consulted items, only attribute pool $\mathcal{P}^{(t+1)}_{x}$ removes $\hat{a}_x$ because user has different preference on attributes in different slots (e.g.,~\textit{white} shirt but \textit{black} pants); (ii)~if $\hat{a}_x$ is accepted, $\mathcal{\check{A}}^{(t+1)}_x$ in short-term context is updated by $\mathcal{\check{A}}^{(t)}_x \cup \{\hat{a}_x\}$; (iii) if $\hat{a}_x$ is explicitly rejected, it only happens when user strongly dislikes this attribute. So $\hat{a}_x$ will be removed from all attributes candidates pools, and items associated with $\hat{a}_x$ will be removed from all item candidate pools as well.
    \item \textbf{Update $\mathbb{\bar{S}}^{(t+1)}_u$:} it is updated by appending a new result id for round $t$, result in a t-sized list.
    \item \textbf{Update slots $\mathcal{X}^{(t+1)}$:} as~\Cref{sec:setting} described, if items accepted, we remove the corresponded slots from $\mathcal{X}^{(t)}$, and create new slots to keep the size as $K$. For a new slot $x'$, the short-term contexts is $(\texttt{[MASK]}, \emptyset)$, and candidate pools are the union sets of previous candidate pools to excluded items or attributes that user strongly dislikes.
\end{itemize}

\subsection{Rewards: Two-Level Reward Definitions}
\label{sec:reward}

We define two-level rewards for these multiple agents. \textbf{(1)~Low-level rewards} are for $\pi_I$ and $\pi_A$,  i.e.,~to make item recommendations and question posting more accurate online. At round $t$, for each slot $x$ the reward $r_x^{I}=1$ if $\pi_I$ hits target item, otherwise 0. Reward $r_x^{A}$ for $\pi_A$ is similar. 
\textbf{(2)~High-level rewards} are for the conversation management agent $\pi_M$ reflecting the quality of a whole conversation. The reward $r^M$ is 0 unless the conversation ends, where we calculate $r^M$ using one of the final bundle metrics (e.g.,~F1 score, accuracy). 

\section{Model Architecture}
\label{sec:model}

Under this framework, we propose a unified model, Bundle BERT (\textsc{Bunt}). In this section, we first describe the architecture of \textsc{Bunt}, then we describe how to train \textsc{Bunt} with offline pre-training and online fine-tuning. 

\textsc{Bunt} is an encoder-decoder framework with multi-type inputs and multi-type outputs to handle user modeling, consultation, and feedback handling. The encoder-decoder framework is commonly used in traditional bundle recommendation tasks~\cite{Bai2019PersonalizedBL, Hu2019Sets2SetsLF, Chen2019POGPO}. We use a self-attentive architecture for three reasons: (1)~Self-attentive models have already been proven as an effective representation encoder and accurate decoder in recommendation tasks~\cite{Kang2018SelfAttentiveSR, Sun2019BERT4RecSR, Chen2019POGPO, he2021locker, Li2021PersonalizedTF}; (2)~RNN~\cite{Cho2014OnTP}-based model inputs have to be ``ordered'', while self-attentive model discards unnecessary order information to reflect the unordered property in bundles; (3)~A self-attentive model can be naturally used in \emph{cloze} tasks (e.g.,~BERT~\cite{Devlin2019BERTPO}), which is suitable for predicting unknown items or attributes in slots.

\begin{figure*}[t]
  \centering
  \includegraphics[width=0.95\linewidth]{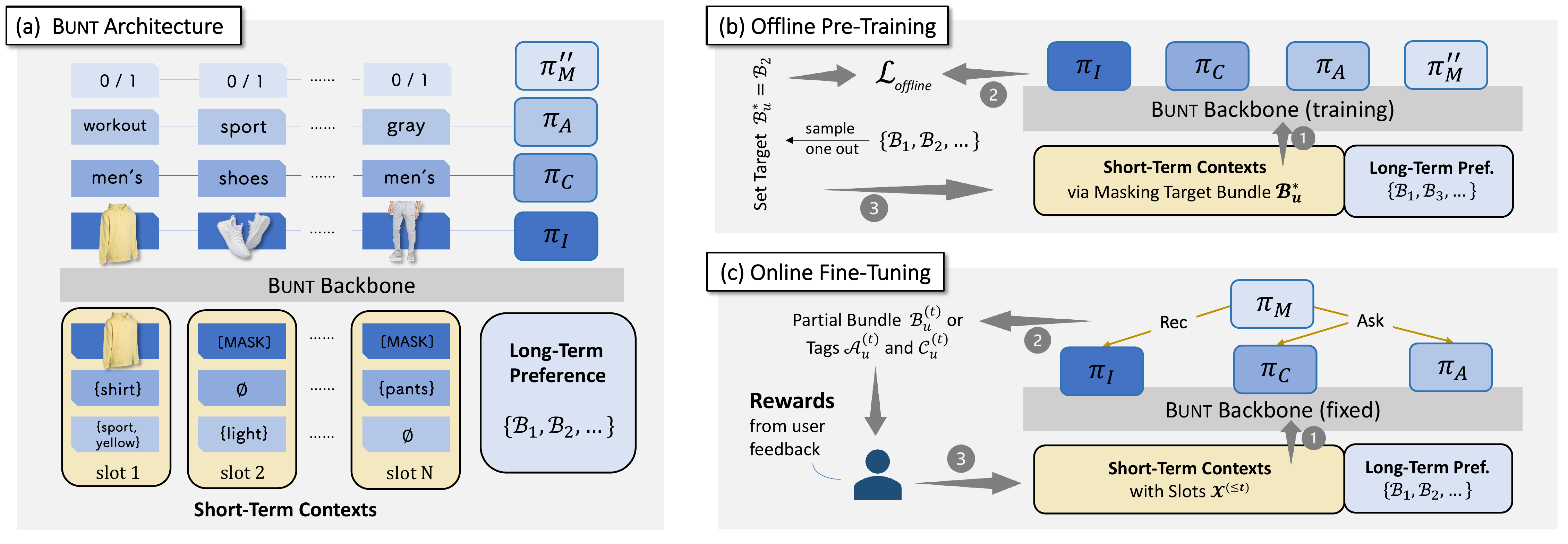}
  \caption{(a)~\textsc{Bunt} architecture illustration. Bunt is a Bert-like model which encodes long-term preference and short-term contexts to infer masked items, categories and attributes per slot $x\in \mathcal{X}^{(t)}$. In this example, $\mathcal{X}^{(t)}=\{2,N\}$ because the related items are still unknown (i.e.,with~\texttt{[MASK]}), and $\mathcal{X}^{(\le t)}=\{1,\dots,N\}$. We define long-term preference and short-term contexts in~\Cref{sec:state}. (b)~\textsc{Bunt} offline pre-training diagram, where \textcircled{\footnotesize{1}} denotes user modeling, \textcircled{\footnotesize{2}} mimics the consultation step, \textcircled{\footnotesize{3}} mimics the feedback handling step, but instead of updating the conversation state at the next round, offline training simply re-masks the target bundle to generate the next masked bundle as \textsc{Bunt} inputs. (c)~\textsc{Bunt} online training diagram, where \textcircled{\footnotesize{1}} is user modeling, \textcircled{\footnotesize{2}} is the consultation step to generate partial bundle $\mathcal{B}^{(t)}_u$ or attributes $\mathcal{A}^{(t)}_u$ and categories $\mathcal{C}^{(t)}_u$, \textcircled{\footnotesize{3}} is the feedback handling step to update short-term contexts. We describe steps \textcircled{\footnotesize{1}}-\textcircled{\footnotesize{3}} in~\Cref{sec:framework,sec:model}, where we keep $\pi_A$ and omit the similar policy $\pi_C$, for ease of description.}
  \label{fig:learning}
  \vspace{-1.2em}
\end{figure*}

\subsection{\textsc{Bunt} for Bundle-Aware User Modeling}
\label{sec:bunt-user-modeling}

\subsubsection{Long-Term Preference Representation}
\label{long-term}

We encode user historical interactions $\{\mathcal{B}_1, \mathcal{B}_2, \dots\}$ as user long-term preferences $\mathbf{E}_u$ using hierarchical transformer (\texttt{TRM})~\cite{SelfAtt} encoders:
\begin{align}
\begin{split}
\label{eq:encoder}
    \mathbf{E}_u &= \texttt{TRM}_{\texttt{bundle}}(\{\mathbf{B}_1, \mathbf{B}_{2}, \dots\}),
    \ \ \text{where}\ \mathbf{B}_n = \texttt{AVG}(\texttt{TRM}_{\texttt{item}}(\mathcal{B}_n)),\ n=1,2,\dots.
\end{split}
\end{align}
$\texttt{TRM}_\texttt{bundle}$ is a transformer encoder over the set of bundle-level representations $\{\mathbf{B}_1, \mathbf{B}_{2}, \dots\}$, the output $\mathbf{E}_u\in\mathbb{R}^{N_u\times d}$ represents user long-term preferences, $N_u$ is the number of historical bundles, and $d$ is the hidden size of the $\texttt{TRM}_\texttt{bundle}$ model. The bundle representation $\mathbf{B}_n\in\mathbb{R}^{1\times d}$ is also extracted by a transformer encoder, namely $ \texttt{TRM}_{\texttt{item}}$, over the set of item embeddings in this bundle, then the set of output embeddings from $\texttt{TRM}_{\texttt{item}}$ is aggregated by average pooling \texttt{AVG} as $\textbf{B}_n$. Our two-level transformers contain no positional embeddings since the input representations are unordered.

\subsubsection{Short-Term Contexts Representation}
We describe how to represent short-term contexts $\{(\check{i}_x^{(t)},  \check{{\mathcal{A}}}^{(t)}_x)| x\in\mathcal{X}^{(\le t)}\}$. We feed the contexts into a special embedding layer $\texttt{EMB}$, then obtain two sets of embeddings for items, attributes: 
\begin{equation}
    \label{eq:emb}
    \mathbf{E}^{(t)}_{\mathit{I},u}, \ \mathbf{E}^{(t)}_{\mathit{A},u}  = \texttt{EMB}(\{(\check{i}_x^{(t)},  \check{{\mathcal{A}}}^{(t)}_x)| x\in\mathcal{X}^{(\le t)}\}),
\end{equation}
where $\mathbf{E}^{(t)}_{*, u}\in\mathbb{R}^{|\mathcal{X}^{(\le t)}|\times d}$ denotes item (I) and attribute (A) embeddings. For items, we retrieve embeddings of the accepted item ids (or \texttt{[MASK]} id). For attributes, we retrieve embeddings corresponding to the accepted attribute ids (or \texttt{[PAD]} id) in $\check{\mathcal{A}}_{x}^{(t)}$ for $x\in\mathcal{X}^{(t)}$, then apply average pooling \texttt{AVG} on embeddings to obtain $\mathbf{E}^{(t)}_{\mathit{A}, u}\in\mathbb{R}^{|\mathcal{X}^{(\le t)}|\times d}$. 

\subsubsection{Long- and Short-Term Representation Fusion}
We feed user long-term preferences $\mathbf{E}_u$ and short-term contexts $\mathbf{E}^{(t)}_{*, u}$ into an $L$-layer transformer. For notation simplicity\footnote{It should be exactly represented as $\mathbf{O}^{(t,0)}_{I, u}$; we omit some notations for simplicity of the decoder description below.}, we denote $\mathbf{E}^{(t)}_{I, u}$ as $\mathbf{O}^{0}$ and get the fused representation:
\begin{align}
\begin{split}
    \label{eq:decoding}
    \mathbf{O}^{l} = \texttt{TRM}_l(\widetilde{\mathbf{O}}^{l-1}, \mathbf{E}_u),\ \widetilde{\mathbf{O}}^{l-1} = \texttt{LN}\left(\mathbf{O}^{l-1}\oplus \mathbf{E}^{(t)}_{A, u}\mathbf{W}^{l-1}\right),
    \ \text{where}\ l=1,\dots,L,
\end{split}
\end{align}
where $\texttt{TRM}_l$ is the $l^\text{th}$ transformer layer with cross attention~\cite{SelfAtt}, $\mathbf{W}^{l-1}$ $\in \mathbb{R}^{d\times d}$ is a learnable projection matrix at layer $l$-1 for attribute representation.  $\oplus$ is element-wise addition and $\texttt{LN}$ denotes LayerNorm~\cite{Vaswani2017Attention} for training stabilization. We incorporate the attribute feature $\mathbf{E}^{(t)}_{A,u}$ before each transformer layer in order to incorporate multi-resolution levels, which is effective in transformer-based recommender models~\cite{Li2020MRIFMI}. Thus for the output representation $\mathbf{O}^L\in\mathbb{R}^{|\mathcal{X}^{(\le t)}|\times d}$, each row  $\mathbf{O}^L_x$ ($x \in \mathcal{X}^{(\le t)}$) contains contextual information from slots in conversation contexts. We treat $\mathbf{O}^L$ and candidate pools $\mathcal{I}_x^{(t)}, \mathcal{P}_x^{(t)}$ for all slots $x\in \mathcal{X}^{(\le t)}$ as the encoded state $\mathbf{S}_u^{(t)}$.
Moreover, for the additional conversation records $\mathbb{\bar{S}}_u^{(t)}$ introduced in~\Cref{sec:state}, we encode it as a vector $\mathbf{\bar{S}}_u^{(t)}$ by using result id embeddings and average pooling.

\subsection{\textsc{Bunt} for Bundle-Aware Consultation}
\label{sec:bunt-consultation}

For consultation step, we feed the encoded state into multiple policy networks to get outputs for each slot $x \in \mathcal{X}^{(t)}$:
\begin{equation}
\label{eq:policy}
        \left\{  
    \begin{aligned}
        &  P_M(a \mid \mathbf{\bar{S}}_u^{(t)}, \mathbf{O}_x^L) = \beta\cdot\pi_M' (a\mid\mathbf{\bar{S}}_u^{(t)}) + (1-\beta)\cdot \pi_M''(a\mid\mathbf{O}_x^L), \quad \text{where}\ a\in{\{0,1\}}, &\small\text{(Conv. Management)} \\\
        & P_I(a \mid \mathbf{O}_x^L) = \pi_I(a \mid \mathbf{O}_x^L), \quad \text{where}\ a\in \mathcal{I}^{(t)}_x, &\small\text{(Bundle Generation)} \\\
        & P_A(a \mid \mathbf{O}_x^L) = \pi_A(a \mid \mathbf{O}_x^L), \quad \text{where}\ a\in \mathcal{P}^{(t)}_x. &\small\text{(Attribute Consultation)}
    \end{aligned}
    \right.
\end{equation}
$P_*$ represents the probability. Policy network $\pi_M$ is linearly combined by two sub models $\pi_M'$ and $\pi_M''$ for state $\mathbf{\bar{S}}_u^{(t)}$ and $\mathbf{O}_x^L$ respectively, $\beta$ is a gating weight\footnote{$\beta$ is predicted by an MLP model with sigmoid function and using concatenated $\bar{S}_u^{(t)}, \mathbf{O}_x^L$ as input.}. $\pi_M'$, $\pi_M''$,
$\pi_I$ and $\pi_A$ are MLP~\cite{MLP} models with ReLU~\cite{Nair2010RectifiedLU} activation and softmax layer. We use $\pi_I$ or $\pi_A$ to infer the masked items or attributes in slot $x$. In inference stage, we take the actions with the highest probability to decide recommending or asking, to construct the consulted (partial) bundle $\mathcal{B}^{(t)}_u$ or questions on attributes $\mathcal{A}^{(t)}_u$. Compared with other individual-item MCR models, the contextual information stored in different slots matters in bundle recommendation, so it is natural to share the state encoded from different slots for both recommendation and question predictions in a unified self-attentive architecture. 

\begin{algorithm}[t]
\small
\caption{\textbf{\textsc{Bunt} Offline Pre-Training}}\label{alg:offline}
\begin{algorithmic}[1]
\Require historical user bundle interactions $\mathcal{D}$, masking ratio $\rho$, \textsc{Bunt} (including $\pi''_M$, $\pi_I$, $\pi_A$) parameters $\Theta$, slot size K;
\Ensure \textsc{Bunt} parameters $\Theta$ after pre-training;
\While {not meet training termination criterion}
\State Sample a user $u \in \mathcal{U}$, get historical bundles $\{\mathcal{B}_1, \dots, \mathcal{B}_{N_u}\}$ from $\mathcal{D}$; sample a historical bundle as target bundle, e.g.,~$\mathcal{B}_n$;
\State Get $\mathbf{E}_u \gets$~\Cref{eq:encoder} with input $\{\mathcal{B}_1, \dots, \mathcal{B}_{N_u}\}\setminus\{\mathcal{B}_n\}$;
\State Sample $l$ items in $\mathcal{B}_n$ as $\mathcal{B}^l_n$; \textcolor{gray}{\Comment{Mimic partial bundle. W.l.o.g.,~assume $|B_n| > K$}}
\State Sample $k\in[1,K]$, then mask $k$ items in $B_n^l$, set the masked positions as slots $\mathcal{X}$;
\State Retrieve attributes for all the items in $\mathcal{B}_n^l$ and mask attributes with probability $\rho$; \textcolor{gray}{\Comment{Mimic short-term contexts}}
\State Predict the distributions of masked items, attributes and conversation management in slot $x\in\mathcal{X}$ via~\Cref{eq:policy};
\State Compute loss $\mathcal{L}_\mathit{offline}$ with~\Cref{eq:rec,eq:ask,eq:conv}; update $\Theta$ using gradient-related optimizer (e.g.,~\cite{Kingma2015AdamAM}).
\EndWhile
\end{algorithmic}
\end{algorithm}

\subsection{Offline Pre-Training}
\label{sec:offline}

Due to the large action spaces of items and attributes, it is difficult to directly train agents from scratch. Thus, we first pre-train the \textsc{Bunt} model on collected offline user-bundle interactions. The core idea of pre-training is to mimic model inputs and outputs in the process of Bundle MCR, which can be treated as multiple \textit{cloze} (i.e.,~``fill the slot'') tasks given a few accepted items and attributes to infer the masked items and attributes. 

\subsubsection{Multi-Task Loss}
\textsc{Bunt} offline training is based on a multi-task loss for recommendation and question asking simultaneously, i.e.,~$
    \mathcal{L}_\mathit{offline} = \mathcal{L}_\mathit{rec} + \lambda \mathcal{L}_\mathit{ask}$,
where $\lambda$ is a trade-off hyper-parameter to balance the importance of these two losses in offline pre-training. We treat item prediction as a multi-class classification task for masked slots $\mathcal{X}^{(t)}$:
\begin{equation}
    \label{eq:rec}
    \mathcal{L}_\mathit{rec} = - \sum_{x\in\mathcal{X}^{(t)}} \sum_{i\in\mathcal{I}^{(t)}_x} y_i\log P_I({i \mid {\mathbf{O}}^L_{x}}),
\end{equation}
where $y_i$ is the binary label (0 or 1) for item $i$. Meanwhile, attribute predictions are formulated as multi-label classification tasks. We use a weighted cross-entropy loss function considering the imbalance of labels to prevent the model from only predicting popular attributes. 
The loss function of attribute predictions is:
\begin{equation}
    \label{eq:ask}
    \mathcal{L}_\mathit{ask} = - \sum_{x\in\mathcal{X}^{(t)}} \sum_{a\in\mathcal{P}^{(t)}_x} w_a\cdot y_a\log P_A({a \mid {\mathbf{O}}^L_{x}}),
\end{equation}
where $w_a$ is a balance weight of attribute $a$ following~\cite{King2001LogisticRI}, and note that multiple $y_a$ can be 1 for multi-label classification. Furthermore, we pre-train part of conversational manager, i.e.,~$\pi''_M$, to decide whether to recommend or ask:
\begin{equation}
    \label{eq:conv}
    \mathcal{L}_\mathit{conv}=-\sum_{x\in\mathcal{X}^{(t)}}\mathbb{I}(l_x\not=-1)\cdot \log \pi''_M(l_x \mid \mathbf{O}_x^L).
\end{equation}
For slot $x$, as long as item agent $\pi_I$ hits the target item, $l_x$ is set as 1; otherwise, if the attribute agent hits the target, $l_x$ is 0. $l_x$ is set as -1 when no agents make successful predictions.
We denote $\mathcal{L}_\mathit{ask} = \mathcal{L}_\mathit{cate} + \mathcal{L}_\mathit{attr} + \mathcal{L}_\mathit{conv}$.

\subsubsection{Training Details} \Cref{fig:learning}b illustrates \textsc{Bunt} offline training. We pre-train \textsc{Bunt} on offline user-bundle interactions, to obtain the basic knowledge to predict the following items or attributes given historical bundle interactions and conversational information. The training details are in~\Cref{alg:offline}. 

\subsection{Online Fine-Tuning}
\label{sec:online}
\Cref{fig:learning}c shows the online-training diagram,
where we fine-tune \textsc{Bunt} agents during interactions with (real or simulated) users. Our core idea is fixing \textsc{Bunt} backbone parameters, fine-tune agents $\pi_I$, $\pi_A$ and $\pi_M$ in a Bundle MCR environment to update related parameters and improve the accuracy after interacting with users. The online fine-tuning details are in~\Cref{alg:online}. We omit the details of RL value networks like~\cite{Lei2020InteractivePR}. 

\begin{algorithm}[t]
\small
\caption{\textbf{Online \textsc{Bunt} Fine-Tuning}}\label{alg:online}
\begin{algorithmic}[1]
\Require trainable \textsc{Bunt} parameters $\Theta_I$, $\Theta_A$ and $\Theta_M$ for three networks $\pi_I$, $\pi_A$ and $\pi_M$, empty buffer $\mathbf{M}_M$, $\mathbf{M}_I$ and $\mathbf{M}_A$;
\Ensure \textsc{Bunt} policy networks parameters $\Theta_I$, $\Theta_A$ and $\Theta_M$;
\For {episode $e=1,2,\dots$}
\State Sample a user $u$, get target bundle  $\mathcal{B}^*_u$; initialize $\mathcal{\check{B}}_u \gets \emptyset$ for recording all the accepted items;
\For{conversation round $t = 1, 2, \dots, T$}
    \State Get conversation states $\mathbf{S}^{(t)}_u$ and $\mathbf{\bar{S}}^{(t)}_u$ via~\Cref{sec:bunt-user-modeling}; get slots $\mathcal{X}^{(t)}$ via~\Cref{sec:transition}; \textcolor{gray}{\Comment{1. user modeling}}
    \State Sample action $a_M$ from $\{0,1\}$ using $\pi_{M}$ via~\Cref{sec:bunt-consultation} ; \textcolor{gray}{\Comment{2. consultation}}
    \If{$a_M$ == 1} 
        \State Use $\mathbf{O}^{L}$ from $\mathbf{S}^{(t)}_u$ to generate a partial bundle $\mathcal{B}^{(t)}_u$ using $\pi_I$ via~\Cref{sec:bunt-consultation}; \textcolor{gray}{\Comment{2.1 recommending}}
        \State Update conversation states $\mathbf{S}^{(t+1)}_u$ and $\mathbf{\bar{S}}^{(t+1)}_u$ via~\Cref{sec:transition,sec:bunt-user-modeling}; get $\mathbf{\tilde{O}}^L$ from $\mathbf{S}^{(t+1)}_u$; \textcolor{gray}{\Comment{3. feedback handling}}
        \State Add $\{(\mathbf{O}^L_x, \mathbf{\tilde{O}}^L_x, \hat{i}_x, r_x^I)\mid x\in\mathcal{X}^{(t)}\}$ to $\mathbf{M}_I$, calculating $r_x^I$ via~\Cref{sec:reward}; \textcolor{gray}{\Comment{i.e., (state, next\_state, action, reward)}}
        \State Add accepted items into $\mathcal{\check{B}}_u$; 
    \ElsIf{$a_M$ == 0} 
        \State Use $\mathbf{O}^{L}$ from $\mathbf{S}^{(t)}_u$ to generate questions on attributes $\mathcal{A}^{(t)}_u$ using $\pi_A$ via~\Cref{sec:bunt-consultation}; \textcolor{gray}{\Comment{2.1 asking}}
        \State Update conversation states $\mathbf{S}^{(t+1)}_u$ and $\mathbf{\bar{S}}^{(t+1)}_u$ via~\Cref{sec:transition,sec:bunt-user-modeling}; get $\mathbf{\tilde{O}}^L$ from $\mathbf{S}^{(t+1)}_u$; \textcolor{gray}{\Comment{3. feedback handling}}
        \State Add $\{(\mathbf{O}^L_x, \mathbf{\tilde{O}}^L_x, \hat{a}_x, r_x^A)\mid x\in\mathcal{X}^{(t)}\}$ to $\mathbf{M}_A$, calculating $r_x^A$ via~\Cref{sec:reward}; \textcolor{gray}{\Comment{i.e., (state, next\_state, action, reward)}}
    \EndIf   
    \State  Add  $((\mathbf{O}^L, \mathbf{\bar{S}}^{(t)}_u), (\mathbf{\tilde{O}}^L, \mathbf{\bar{S}}^{(t+1)}_u ), a_M, r^M)$ to $\mathbf{M}_M$, calculating $r^M$ via~\Cref{sec:reward}; \textcolor{gray}{~\Comment{i.e.,~(state, next\_state, action, reward)}} 
    
    \If{$\mathcal{\check{B}}_u=\mathcal{B}^*_u$ or $t=T$}
        \State Current conversation terminates;
    \EndIf 
\EndFor
\If{$\mathbf{M}_k$  ($k=\{M,I,A\}$) meets pre-defined buffer training criterion (e.g.,~buffer size)}
    \State Update $\Theta_k$ using $\mathbf{M}_k$ with RL methods (e.g.,~DQN~\cite{DQN}, PPO~\cite{Schulman2017ProximalPO}); Then reset $\mathbf{M}_k$; \textcolor{gray}{\Comment{policy learning}}
\EndIf 
\EndFor
\end{algorithmic}
\end{algorithm}

\section{Experiments}

\subsection{Evaluation Protocol and Metrics}
\label{sec:metric}
Following~\cite{Deng2021BuildYO, Kang2018SelfAttentiveSR, Lei2020InteractivePR}, we conduct a \emph{leave-one-out} data split (i.e.~for each user, randomly select $N$-1 bundles for offline training, the last bundles for online training, validation and testing respectively in a ratio of 6:2:2). We choose the multi-label precision, recall, F1, and accuracy, defined in~\cite{Zhang2014ARO} to measure the quality of the generated bundle.

\subsection{Datasets}

We extend four datasets (see statistics in~\Cref{tab:data}) for Bundle MCR.\footnote{We cannot use other bundle datasets such as \textit{Youshu} or \textit{NetEase} because they do not provide item attributes or categories information.} 
\textbf{(1)~Steam}: This dataset collects user interactions with game bundles in~\cite{Pathak2017GeneratingAP} from the Steam\footnote{\url{https://store.steampowered.com}} platform. We use item \emph{tags} as \emph{attributes} in Bundle MCR and item \emph{genres} as \emph{categories} and discard users with fewer than two bundles according to our evaluation protocol.
\textbf{(2)~MovieLens}: This dataset is a benchmark dataset~\cite{Harper2015TheMD} for collaborative filtering tasks. We use the ML-10M version by treating movies rated with the same timestamps (second-level granularity) as a bundle. We treat provided \textit{genres} as \textit{categories}, \textit{tags} as \textit{attributes} in Bundle MCR.
\textbf{(3)~Clothing}: This dataset is collected in~\cite{McAuley2015ImageBasedRO} from Amazon\footnote{\url{https://www.amazon.com}} e-commerce platform; we use the 
subcategory \textit{clothing}. We treat co-purchased items as a bundle by timestamp. We use item \emph{categories} in the metadata as \emph{categories} in Bundle MCR, and \emph{style} in item reviews (\emph{style} is a dictionary of the product metadata, e.g.,~``format'' is ``hardcover'', we use ``hardcover'') as \emph{attributes}. For MovieLens and Clothing, bundles are grouped by timestamp thus noisy. To improve data quality, we filter out users and items that appear no more than three times. 
\textbf{(4)~iFashion} is an outfit dataset with user interactions~\cite{Chen2019POGPO}. Similar to~\cite{Wang2021LearningIB}, we pre-process iFashion as a 10-core dataset to ensure data quality. We use \emph{categories} features from iFashion metadata, and tokenize the \textit{title} as \textit{attributes} in Bundle MCR. 

\subsection{Baselines}

We introduce three groups of recommendation baselines to evaluate Bundle MCR and \textsc{Bunt} (we call our full proposed \textsc{Bunt} in~\Cref{sec:model} as \textsc{Bunt}-Learn). More technical details of baselines can be found in~\Cref{sec:related}. 
\subsubsection{Traditional bundle recommenders}  
\textbf{Freq} uses the most frequent bundle as the predicted bundle. \textbf{BBPR~\cite{Pathak2017GeneratingAP}:} considering the infeasible time cost of cold bundle generation in BBPR, we use BBPR to rank existing bundles. \textbf{BGN~\cite{Bai2019PersonalizedBL}} adopts an encoder-decoder~\cite{Sutskever2014SequenceTS} architecture to encode user historical interactions and generate a sequence of items as a bundle. We use the top-1 bundle in BGN generated bundle list as the result. \textbf{{PoG~\cite{Chen2019POGPO}}} is a transformer-based~\cite{SelfAtt} encoder-decoder model to generate personalized outfits. We use it for general bundle recommendation. \textbf{{BYOB~\cite{Deng2021BuildYO}}} is the most recent bundle generator using reinforcement learning methods.  

\begin{table}[]
\small
\caption{Data Statistics, where \# denotes quantity number, \emph{U} denotes user, \emph{I} denotes item, \emph{B} denotes bundle, \emph{C} denotes category and \emph{A} denotes attributes. \emph{B/U} represents the number of bundles per user, \emph{B size} represents the average number of items per bundle.}
\label{tab:data}
\setlength{\tabcolsep}{5pt}
\begin{tabular}{lcccccccccccc}
\toprule
  Dataset & \#U & \#I & \#B & \#C & \#A & \#Inter & B/U & {B Size} & {\#Offline} & {\#Online} & {\#Valid} & {\#Test} \\ \midrule
  Steam            & 13,260       & 2,819        & 229          & 21           & 327          & 261,241          & 2.95         & 5.76            & 13,260             & 7,956             & 2,652            & 2,652           \\
  MovieLens        & 46,322       & 5,899        & 851,361      & 19           & 190          & 3,997,583        & 27.81        & 3.11            & 46,322             & 27,793            & 9,264            & 9,265           \\
  Clothing         & 19,065       & 25,408       & 79,610       & 668          & 4,027        & 285,391          & 5.03         & 3.17            & 19,065             & 11,439            & 3,813            & 3,813           \\ 
    iFashion         & 340,762      & 68,921       & 5,593,387    & 61           & 4,264        & 21,552,716       & 16.41        & 3.79            & 340,762            & 204,457           & 68,152           & 68,153          \\ \bottomrule
\end{tabular}
\end{table}

\subsubsection{Adopted individual recommenders for Bundle MCR}\textbf{FM-All} is an FM~\cite{Rendle2010FactorizationM} variant used in MCR frameworks~\cite{Lei2020EstimationActionReflectionTD, Lei2020InteractivePR}, ``All'' means this model in Bundle MCR only recommends top-$K$ items per round without asking any questions. \textbf{FM-Learn} follows the item predictions in FM-All, but use other pre-trained agents in \textsc{Bunt} for conversation management and question posting. \textbf{EAR~\cite{Lei2020EstimationActionReflectionTD}} and \textbf{SCPR~\cite{Lei2020InteractivePR}} are popular Individual MCR frameworks based on FM. We keep the core ideas of estimation-action-reflection in our EAR and asking attributes by path reasoning in our SCPR, and change the names to EAR* and SCPR* because some implemented components are changed for adapting into Bundle MCR. We do not use recent {UNICORN~\cite{UNICORN}} or {MIMCR~\cite{MIMCR}}, because the unified action space in UNICORN is incompatible with Bundle MCR to generate multiple items or attributes per round; main contributions of MIMCR are based on the multiple choice questions setting, which is incompatible with Bundle MCR.

\subsubsection{Simple bundle recommenders for Bundle MCR}  \textbf{\textsc{Bunt}-One-Shot} uses \textsc{Bunt} in traditional bundle recommendation following the inference of PoG~\cite{Chen2019POGPO}. \textbf{\{BYOB, BGN, \textsc{Bunt}\}-All} models are simple bundle recommender implementations in Bundle MCR, only recommending top-$K$ items per round without asking any questions. 

\subsection{Experimental Setup}
\subsubsection{Training Details}
Our training phases are two-stage\footnote{\change{More details of metric definitons, data processing, \textsc{Bunt} implementation and human evaluation setup are in \url{https://github.com/AaronHeee/Bundle-MCR}.}}: (1)~in offline pre-training, we follow~\Cref{alg:offline} to implement and train our \textsc{Bunt} model with \texttt{PyTorch}. The number of transformer layers and heads are searched from \{1,2,4\}, $d=32$, $K=2$, $\lambda=0.1$ and masking ratio $\rho=0.5$. We use an Adam~\cite{Kingma2015AdamAM} optimizer with initial learning rate 1e-3 for all datasets with batch size 32. The maximum bundle size is set as $20$.
(2)~In online fine-tuning, we implement~\Cref{alg:online} using \texttt{OpenAI} \texttt{Stable-Baselines} RL training code. We reuse Proximal Policy Optimization~\cite{Schulman2017ProximalPO} (PPO) in \texttt{Stable-Baselines}\footnote{\url{https://stable-baselines3.readthedocs.io}} to train four agents ($\pi_M$, $\pi_I$, $\pi_C$, $\pi_A$) jointly ($\pi_C$ is category policy, similar to $\pi_A$) using Adam optimizer with lr=1e-3. Other hyper-parameters follow the default settings in \texttt{Stable-Baselines}. We re-run all experiments three times with different random seeds and report the average performance and related standard errors.

\begin{table}[]
\small
\caption{\textsc{Bunt} and other individual conversational recommendation methods that are adopted for bundle settings. The best is \textbf{bold}.}
\setlength{\tabcolsep}{4pt}
\renewcommand{\arraystretch}{1}
\begin{tabular}{clcccccccc}
\toprule
                                                                                    &                   & \multicolumn{4}{c}{Steam}                             & \multicolumn{4}{c}{MovieLens}                          \\ \cmidrule(l){3-6} \cmidrule(l){7-10} 
Group                                                                               &   Method        & Precision   & Recall      & F1          & Accuracy    & Precision   & Recall      & F1          & Accuracy    \\ \midrule
\multirow{4}{*}{\begin{tabular}[c]{@{}c@{}}Individual\\ Rec.\\ Model\end{tabular}}  & (a) FM-All        & .149±.001 & .611±.004 & .239±.001 & .138±.001 & .019±.001 & .087±.002 & .031±.001 & .017±.001 \\
                                                                                    & (b) FM-Learn      & .269±.019 & .664±.018 & .382±.016 & .239±.001 & .038±.002 & .096±.008 & .055±.005 & .031±.002 \\
                                                                                    & (c) EAR*          & .186±.034 & .592±.025 & .282±.041 & .166±.031 & .036±.003 & .099±.009 & .053±.005 & .029±.003 \\
                                                                                    & (d) SCPR*         & .173±.009 & .544±.043 & .262±.008 & .151±.006 & .044±.012 & .110±.007 & .063±.009 & .032±.006 \\ \midrule
\multirow{5}{*}{\begin{tabular}[c]{@{}c@{}}Bundle \\ Rec.\\ Model\end{tabular}}     & (e) \textsc{Bunt}-One-Shot & .456±.006 & .452±.007 & .454±.007 & .450±.006 & .075±.007 & .093±.006 & .083±.007 & .061±.005 \\
                                                                                    & (f) BYOB-All      & .328±.046 & .799±.023 & .463±.047 & .323±.047 & .020±.001 & .113±.007 & .034±.002 & .018±.001 \\
                                                                                    & (g) BGN-All       & .568±.019 & .919±.007 & .702±.013 & .567±.019 & .073±.005 & .216±.006 & .109±.006 & .070±.006 \\
                                                                                    & (h) \textsc{Bunt}-All      & .633±.012 & .927±.002 & .752±.008 & .632±.012 &  .100±.004 & .289±.004 & .149±.004 & .096±.003  \\  \cmidrule(l){2-10}
                                                                                    & \textbf{(i) \textsc{Bunt}-Learn}    & \textbf{.737±.003} & \textbf{.928±.015} & \textbf{.822±.006} & \textbf{.727±.006} &  \textbf{.251±.015} & \textbf{.302±.013} & \textbf{.275±.013} & \textbf{.181±.008} \\ \bottomrule \toprule 
                                                                                    &                   & \multicolumn{4}{c}{Clothing}                         & \multicolumn{4}{c}{iFashion}                          \\ \cmidrule(l){3-6} \cmidrule(l){7-10} 
Group                                                                               &   Method        & Precision   & Recall      & F1          & Accuracy    & Precision   & Recall      & F1          & Accuracy    \\ \midrule
\multirow{4}{*}{\begin{tabular}[c]{@{}c@{}}Individual \\ Rec.\\ Model\end{tabular}} & (a) FM-All        & .003±.001 & .013±.001 & .005±.001 & .003±.001 & .006±.001 & .026±.001 & .010±.001 & .005±.001 \\
                                                                                    & (b) FM-Learn      &  .006±.001 & .010±.003& .008±.002 & .004±.001  &  .008±.003  & 028±.002  & .012±.002  &  .006±.002           \\
                                                                                    & (c) EAR*          &  .011±.003 & .022±.002 & .014±.003 & .008±.002 & .017±.003 & .026±.001 & .020±.002 & .010±.001 \\
                                                                                    & (d) SCPR*         &  .013±.006 & .028±.004 &  .018±.003 & .009±.005  &   .014±.005  &  .032±.003  & .019±.004  &.010±.002             \\ \midrule
\multirow{5}{*}{\begin{tabular}[c]{@{}c@{}}Bundle \\ Rec.\\ Model\end{tabular}}     & (e) \textsc{Bunt}-One-Shot &  .006±.001 & .005±.001 & .005±.001 & .005±.001 &  .008±.002 & .007±.001 & .007±.001 & .005±.002 \\
                                                                                    & (f) BYOB-All      & .002±.001 & .010±.001 & .003±.001 & .002±.001 &  .005±.001 & .023±.001 & .008±.001 & .004±.001  \\
                                                                                    & (g) BGN-All       &.009±.001 & .023±.002 & .013±.001 & .008±.001 &  .011±.001 & .032±.002 & .016±.002 & .010±.001 \\
                                                                                    & (h) \textsc{Bunt}-All      &  .008±.001 & .023±.002 & .012±.002 & .008±.001 & .014±.001 & \textbf{.043±.001} & .021±.001 & .014±.001  \\  \cmidrule(l){2-10}
                                                                                    & \textbf{(i) \textsc{Bunt}-Learn }   &  \textbf{.019±.003} & \textbf{.026±.008} & \textbf{.021±.005} & \textbf{.015±.004} & \textbf{.020±.001} & .035±.003 & \textbf{.025±.001} & \textbf{.017±.001} \\ \bottomrule
\end{tabular}
\label{tab:main-table}
\end{table}

\subsubsection{User Simulator Setup} 
\label{sec:simulator}
Due to the difficulty and cost of interacting with real users, we mainly evaluate our frameworks with user simulators, similar to previous works~\cite{Lei2020EstimationActionReflectionTD, Lei2020InteractivePR, UNICORN, MIMCR}. We simulate a user with a target bundle $\mathcal{B}^*$ in mind, which is sampled from our online dataset. To mimic real user behavior, the user simulator \emph{accepts} system-provided items which agree with target bundle $\mathcal{B}^*$; \emph{accepts} categories and attributes that agree with the potential target items in the current slot.\footnote{Given a slot $x$, initially all items in $\mathcal{B}^*$ are potential items, but some items are removed with the acceptance of items, categories and attributes in slot $x$.}  The user simulator only explicitly \emph{rejects} categories or attributes that are not associated with any items in $\mathcal{B}^*$. In other cases, the user simulator \emph{ignores} items, categories and attributes provided by system. The user simulator is able to terminate conversations when all items in $\mathcal{B}^*$ have been recommended. Otherwise, the system ends conversations after $t=T$ conversation rounds. We set the maximum conversation rounds $T$ to 10 in our experiments.

\subsection{Main Performance of Bundle MCR and \textsc{Bunt}-Learn}
\label{sec:main}

\Cref{tab:main-table} and~\Cref{fig:3} show the main performance of our proposed framework and model architecture compared with other conversational recommendation baselines. 
We make the following observations:

\subsubsection{\textsc{Bunt} Backbone Performance} Though we propose \textsc{Bunt} for the Bundle MCR task, we first show \textsc{Bunt} is competitive in traditional \emph{one-shot} bundle recommendation. \Cref{fig:tradition} shows \textsc{Bunt} outperforms classic bundle recommenders (i.e.,~BBPR) markedly, and is comparable to or sometimes better than  recent bundle generators (e.g.,~BGN, PoG). In this regard, \textsc{Bunt} backbone is shown to learn basic ``bundle recommendation'' knowledge much like other models. 

\subsubsection{Effectiveness of Bundle MCR}
\label{sec:exp_mcr}

We show the effectiveness of Bundle MCR by comparing model~(e) and (i) in~\Cref{tab:main-table}. For example, the accuracy on MovieLens data is improved from 0.061 to 0.181. This indicates even given the same backbone model, introducing a conversational mechanism (Bundle MCR) can collect immediate feedback and improve recommendation performance. Also, we observe the relative improvement on the other three datasets is higher than on Steam. For example, the relative improvement in accuracy is 61.56\% 
on Steam compared to 196.72\% on MovieLens. This shows challenging datasets (e.g.,~sparser, larger item space) can gain more benefit from Bundle MCR, since it allows users to provide feedback during conversations.

\subsubsection{Effectiveness of \textsc{Bunt}-Learn.}
\label{sec:exp_bunt_learn}
We adopted several Individual MCR recommenders (a)-(d) in~\Cref{tab:main-table} into bundle settings, in which the backbone model (FM~\cite{Rendle2010FactorizationM}) recommends top-K items without considering bundle contexts. Compared with these individual MCR recommenders, \emph{\textsc{Bunt}-Learn} achieves the best performance. For example, compared with model (b), where we only replace \textsc{Bunt} backbone with FM, ours improves accuracy from 0.239 to 0.727 on Steam. This shows that directly applying existing individual MCR recommenders in Bundle MCR is not optimal, and also shows the benefits of our \textsc{Bunt} design. Moreover, compared with bundle recommenders only recommending items (models (f)-(h)) ours introduces \emph{question asking} and improves recommendation performances consistently, except for the recall score in iFashion. This is because we should replace recommending with asking, so recall may drop given fewer recommendation rounds (F1-Score is improved still).

\begin{figure}[t]
\begin{subfigure}[b]{0.56\textwidth}
  \centering
  \includegraphics[width=\textwidth]{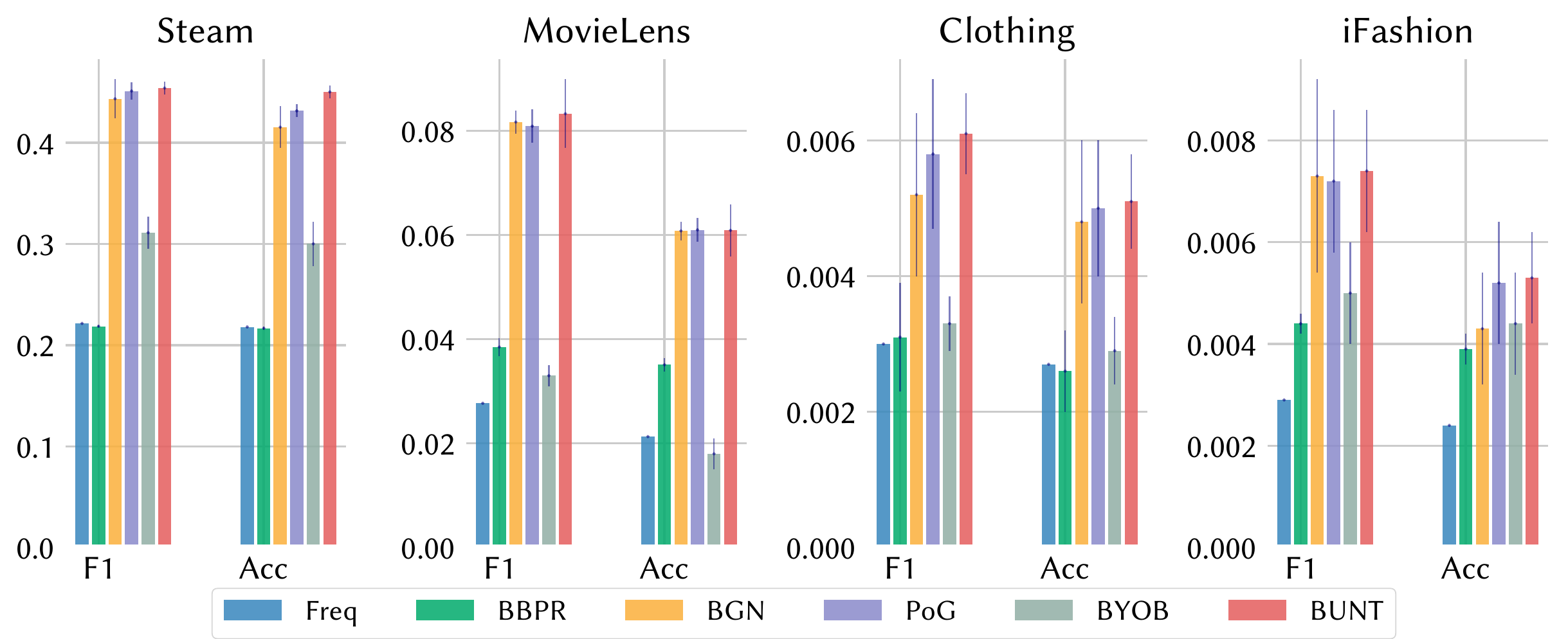}
  \caption{\textsc{Bunt} performance compared with other bundle recommenders in \emph{one-shot} bundle recommendation.}
  \label{fig:tradition}
\end{subfigure}
\hfill
\begin{subfigure}[b]{0.425\textwidth}
     \centering
        \includegraphics[width=\textwidth]{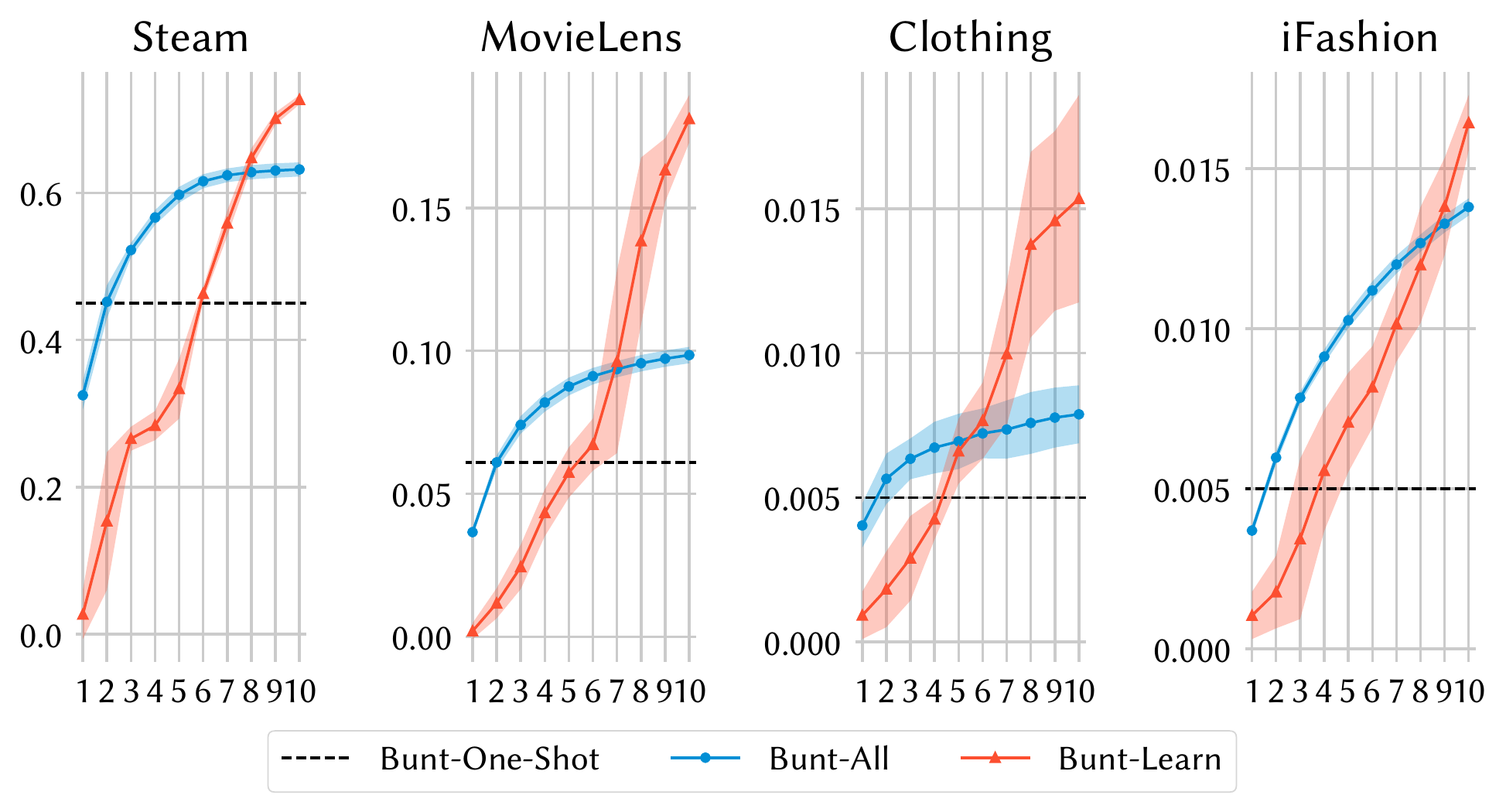}
        \caption{
        Cumulative
        accuracy curve, i.e.,~accuracy  after $t$ rounds, $t\in[1,10]$.}
        \label{fig:curve}
\end{subfigure}
\caption{\change{\textsc{Bunt} performance in \emph{one-shot} setting, and cumulative accuracy curves.}}
\label{fig:3}
\end{figure}

\subsubsection{Accuracy Curve with Conversation Rounds.} The cumulative accuracy curves in~\Cref{fig:curve} show \textsc{Bunt}-All achieves the best results in beginning conversation rounds, then is outperformed by \textsc{Bunt}-Learn. This is because \textsc{Bunt}-Learn requires several rounds to ask questions and elicit preferences. Thus, \textsc{Bunt}-Learn in late rounds can recommend more accurately and surpasses baselines. For example, on MovieLens, \textsc{Bunt}-Learn outperforms the baselines after $t=6$.

\begin{table}[t]
\label{tab:ablation}
\caption{Ablation Studies (F1-score) to evaluate model architecture, fine-tuning (FT) and pre-training (PT).}
\setlength{\tabcolsep}{3pt}
\renewcommand{\arraystretch}{1.1}

\begin{subtable}[h]{0.38\textwidth}
\footnotesize
\begin{tabular}{lcc}
\toprule
Ablation         & Steam & MovieLens \\ \midrule
\textbf{(a) Bunt-Learn}                           &   \textbf{.822±.006}    &     \textbf{.275±.013}      \\ \midrule
(b) w/o Long-term Pref.               & .701±.080      &   .148±.010        \\
(c) w/o Short-term Tags      &  .787±.013     &    .165±.012       \\
(d) w/o Short-term Items  &  .330±.011     &  .084±.009         \\
(e) replace \textsc{Bunt} $\pi_I$ with FM & .382±.016    &   .032±.006        \\ \bottomrule
\end{tabular}
\end{subtable}
\hspace*{\fill}%
\begin{subtable}[h]{0.3\textwidth}
\footnotesize
\begin{tabular}{lcc}
\toprule
Ablation         & Steam & MovieLens \\ \midrule
\textbf{(a) Bunt-Learn }                          &   \textbf{.822±.006}    &     \textbf{.275±.013}      \\ \midrule
(f) w/o FT $\pi_M$               &  .765±.002     &  .210±.002         \\
(g) w/o FT $\pi_I$ &   .817±.003    &   .268±.007        \\
(h) w/o FT $\pi_{\{A,C\}}$      &  .811±.001     &   .257±.003        \\
(i) w/o FT All &   .753±.008    &         .206±.008  \\ \bottomrule
\end{tabular}
\end{subtable}
\hspace*{\fill}%
\begin{subtable}[h]{0.3\textwidth}
\footnotesize
\begin{tabular}{lcc}
\toprule
Ablation                       & Steam & MovieLens \\ \midrule
\textbf{(a) Bunt-Learn  }                         &   \textbf{.822±.006}    &     \textbf{.275±.013}      \\ \midrule
(j) w/o PT $\pi_M$               &  .807±.022     &    .258±.003       \\
(k) w/o PT $\pi_I$ &  .056±.002     &    .008±.001       \\ 
(l) w/o PT $\pi_{\{A,C\}}$      &   .815±.018    &   .171±.002        \\
(m) w/o PT All &   .003±.001    &   .001±.001        \\ \bottomrule
\end{tabular}
\end{subtable}
\end{table}

\subsubsection{\textsc{Bunt}-Learn Component Analysis}
Compared with (a), (b)-(d) show the effectiveness of long-term preference and short-term context encoding; (e) indicates the importance of using bundle-aware models; (f)-(i) show the benefit of online fine-tuning, which helps $\pi_M$ most because conversation management is hard to mimic in offline datasets, and $\pi_M$ with only a binary action space is easier for online learning than other policies; (j)-(m) show pre-training is necessary, especially for item policy, because bundle generation is challenging to directly learn from online interactions with RL. This also indicates the proposed multiple cloze pre-training tasks are suitable for training Bundle MCR effectively.

\subsection{Human Evaluation on Conversation Trajectories}

Considering the cost of deploying real interactive Bundle MCRs, similar to \cite{Xian2021EXACTAEC, Jannach2020EndtoEndLF}, we conduct human evaluation by letting real users rate the generated conversation trajectories from~\Cref{sec:main}. From Steam and MovieLens datasets, we sample 1000 (in total) pairs of conversation trajectories from <\textsc{Bunt}-Learn, SCPR*> or <\textsc{Bunt}-Learn, FM-Learn> (SCPR* and FM-Learn are the best baselines using individual item recommenders). Each pair of conversation trajectories is posted to \change{collect 5 answers} from MTurk\footnote{\url{https://requester.mturk.com/}} workers, who are required to measure the subjective quality by browsing the conversations and selecting the best model from the given pair. We use the answers from high-quality workers who spend more than 30 seconds and the \texttt{LifeTimeAcceptanceRate} is 100\%, and count the majority votes per pair. Lastly, we collected 388 valid results:  <\textsc{Bunt}-Learn, SCPR*> votes are 121:88, and <\textsc{Bunt}-Learn, FM-Learn> votes are 110:69. This result shows the superiority of
\textsc{Bunt}-Learn. Interestingly, the performance gap in human evaluation is not as large as results in simulators (e.g., on Steam, \textsc{Bunt}-Learn accuracy is 3x as FM-Learn). 

\section{Conclusion and Future Work}
\label{sec:conclusion}

In this work, we first extend existing Multi-Round Conversational Recommendation (MCR) settings to bundle recommendation scenarios, which we formulate as an MDP problem with multiple agents. Then, we propose a model architecture, \textsc{Bunt}, to handle bundle contexts in conversations. Lastly, to let \textsc{Bunt} learn bundle knowledge from offline datasets and an online environment, we propose a two-stage training strategy to train our \textsc{Bunt} model with multiple \textit{cloze} tasks and multi-agent reinforcement learning respectively. We show the effectiveness of our model and training strategy on four offline bundle datasets and human evaluation. Since ours is the first work to consider conversation mechanisms in bundle recommendation, many research directions can be explored in the future. In \textsc{Bunt}, our question spaces are about categories and attributes, so how to use \emph{free text} in bundle conversational recommendation is still an open question. Meanwhile, how to explicitly incorporate item relationships (e.g.,~substitutes, complements) in conversational bundle recommendation should be another interesting and challenging task. Moreover, since individual items can be treated as a special bundle, it is interesting to unify existing individual conversational recommenders into conversational bundle recommendation, i.e.,~augmenting conversational agents' abilities without extra cost.


\begin{acks}
\change{We thank the reviewers for their insightful comments, and thank Yisong Miao for precious discussions on this project.}
\end{acks}

\bibliographystyle{ACM-Reference-Format}
\bibliography{7_reference}

\end{document}